\documentclass[preprints,article,accept,moreauthors,pdftex,10pt,a4paper]{mdpi} 
\pdfoutput=1

\firstpage{1} 
\pubvolume{xx}
\issuenum{1}
\articlenumber{1}
\pubyear{2019}
\copyrightyear{2019}
\history{}

\usepackage{xcolor}
\usepackage{mathtools}

\newcommand{\imUnit}{\mathbf{i}}

\newcommand{\fio}[1]{\hat{\mathbb{I}}\left(#1\right)}

\makeatletter
\def\bstr{b}
\def\bfstr{bf}
\def\cstr{c}
\def\fstr{f}
\def\lst{A,B,C,D,d,E,F,G,H,I,J,K,L,M,N,O,P,Q,R,S,T,U,V,W,X,Y,Z,b}

\newcommand{\MkB}[1]{\expandafter\def\csname\bstr#1\endcsname{\mathbb{#1}}}
  \@for\i:=\lst\do{%
    \expandafter\MkB \i     }
    
\newcommand{\MkBF}[1]{\expandafter\def\csname\bfstr#1\endcsname{\mathbf{#1}}}
  \@for\i:=\lst\do{%
    \expandafter\MkBF \i     }

\newcommand{\MkCal}[1]{\expandafter\def\csname\cstr#1\endcsname{\mathcal{#1}}}
  \@for\i:=\lst\do{%
    \expandafter\MkCal \i     }
    
\newcommand{\MkFrak}[1]{\expandafter\def\csname\fstr#1\endcsname{\mathfrak{#1}}}
  \@for\i:=\lst\do{%
    \expandafter\MkFrak \i     }
    
\makeatother

\newcommand{\tio}{\hat{\mathbf{\mathbb{I}}}}

\newcommand{\dom}[1]{\mathop{dom}(#1)}

\newcommand{\range}[1]{\mathop{range}(#1)}
\newcommand{\supp}[1]{\mathop{supp}(#1)}

\newmuskip\pFqmuskip

\newcommand*\pFq[6][8]{%
  \begingroup 
  \pFqmuskip=#1mu\relax
  \mathchardef\normalcomma=\mathcode`,
  \mathcode`\,=\string"8000
  \begingroup\lccode`\~=`\,
  \lowercase{\endgroup\let~}\pFqcomma
  {}_{#2}F_{#3}{\left[\genfrac..{0pt}{}{#4}{#5};#6\right]}%
  \endgroup
}
\newcommand{\pFqcomma}{{\normalcomma}\mskip\pFqmuskip}

\DeclarePairedDelimiterXPP\seq[2]{}{\big(}{\big)}{_{#2}}{#1}

\Title{Operator Ordering and Solution of Pseudo-Evolutionary Equations}

\Author{Nicolas Behr $^{1}$, Giuseppe Dattoli $^{2,*}$ and Ambra Lattanzi $^{2,3}$}

\AuthorNames{Nicolas Behr, Giuseppe Dattoli and Ambra Lattanzi}

\address{%
$^{1}$ \quad Institut de Recherche en Informatique Fondamentale (IRIF), Universit\'{e} Paris-Diderott, F-75013 Paris, France\\
$^{2}$ \quad ENEA -- Frascati Research Center, Via Enrico Fermi 45, 00044 Rome, Italy\\
$^{3}$ \quad H.~Niewodnicza\'{n}ski Institute of Nuclear Physics, Polish Academy of Science, 31-342 Krak\'{o}w, Poland}

\corres{Correspondence: giuseppe.dattoli@enea.it}

\abstract{The solution of pseudo initial value differential equations, either ordinary or partial (including those of fractional nature), requires the development of adequate analytical methods, complementing those well established in the ordinary differential equation setting. A combination of techniques, involving procedures of umbral and of operational nature, has been demonstrated to be a very promising tool in order to approach within a unifying context non-canonical evolution problems. This article covers the extension of this approach to the solution of pseudo-evolutionary equations. We will comment on the explicit formulation of the necessary techniques, which are based on certain time- and operator ordering tools. We will in particular demonstrate how Volterra-Neumann expansions, Feynman-Dyson series and other popular tools can be profitably extended to obtain solutions of fractional differential equations. We apply the method to a number of examples, in which fractional calculus and a certain umbral image calculus play a role of central importance.}



\begin{document}

\section{Introduction}

We discuss in this paper the notion of so-called \emph{pseudo-evolutionary differential equations}, which we define as equations of the Cauchy type,
 \begin{equation}\label{eq:PEO}
\hat{D}_t F(x,t)=\alpha \;\hat{O}_x\,F(x,t)\,\qquad F(x,0)=f(x)\,.
 \end{equation}
Here, $\hat D_t$  is an operator playing the role of the derivative with respect to the variable  $t$, $\alpha$ is a constant, $\hat O_x$ is an operator of differential or pseudo-differential nature (to be specified later in our discussion), and where $f(x)$ is the initial condition.

The development of suitable solution methods, mimicking those exploited for the ordinary Cauchy problem setting, requires the following steps:
\begin{enumerate}
\item Finding an eigenfunction of the $\hat D_t$ operator, such that
\begin{equation}
 \hat{D}_t E(\lambda t)=\lambda E(\lambda t)\,.
 \end{equation}
\item Constructing a \emph{pseudo evolution operator (PEO)} as
\begin{equation}
\hat{U}(t)=E(\alpha \;t\;\hat{O}_x)\,,
\end{equation}
which leads to the solution of~\eqref{eq:PEO} via 
\begin{equation}\label{eq:Fev}
F(x,t)=E(\alpha \;t\;\hat{O}_x)f(x)\,.
\end{equation}
\item Establishing rules that permit  the explicit evaluation of the action of the PEO $\hat{U}(t)$ on the initial function  $f(x)$ in the formal solution~\eqref{eq:Fev} .
\end{enumerate}
Consequently, the precise properties of the eigenfunctions $E$ that arise in the first step of this procedure are of crucial importance in terms of developing explicit calculation techniques. %
It is evident that the strategy we have in mind is that of treating $E$ as a kind of exponential function (the eigenfunction of the ordinary derivative operator) --- in this way, we can recover most of the techniques associated with
operator disentanglement and time-ordering (whenever necessary).

A first study in this direction has been addressed by D.~Babusci and one of the present authors in~\cite{babusci2011umbral}, which has later been specialized in a number of papers for situations in which  $\hat D_t$ is a fractional~\cite{Dattoli_2017} or Laguerre~\cite{dattoli1999hermite} derivative. %
In elaborating this program, we have been faced with some difficulties associated with the
fact that the corresponding eigenfunctions $E$ lack the semi-group property, namely
\begin{equation}
E(x+y)\neq E(x)\;E(y)=E(y)\;E(x)\,.
\end{equation}
This is an additional source of difficulties also when dealing with operator-ordering in pseudo-evolutionary problems. If $\hat{x},\hat{y}$ are non-commuting operators, such that 
\begin{equation}
 \left[\hat{x}\;,\;\hat{y}\right]\neq 0\,,
 \end{equation}
it is well known that even for the exponential function one finds that~\cite{weyl1950theory}
\begin{equation}
 e^{\hat x+\hat y}\neq e^{\hat x}e^{\hat y}\neq e^{\hat y}e^{\hat x}\,.
\end{equation}
Instead, one finds the following types of expansions~\cite{Wilcox_1967, Dattoli_1988}
\begin{equation}
 e^{\hat x+\hat y}
 	=e^{\hat x}e^{\hat y}e^{\hat f_1}\dotsc e^{\hat f_n}\dotsc
	=e^{\hat y}e^{\hat x}e^{\hat g_1}\dotsc e^{\hat
g_n}\dotsc\,,
\end{equation}
where $\hat f_n,\;\hat g_n$ are expressed in terms of chains of commutators involving  $\hat x\;,\;\hat y$, as it happens e.g.\ for the Zassenhaus expansion~\cite{Casas:2012aa}. It is evident that in the special case  $\left[\hat x\;,\;\hat y\right]=0$, the $\hat f_n$ and $\hat g_n$ operator functions vanish, and the semi-group property is restored.

Time-ordering is a further element of complication, which may arise in evolutionary problems (ordinary or pseudo) whenever the operator $\hat O_x$ on the rhs of~\eqref{eq:PEO} is explicitly time-dependent and does not commute with itself at different times.

In this paper we reconsider the operator-ordering problem for pseudo-evolutionary problems in more general terms than those considered in~\cite{babusci2011umbral,Dattoli_2017,dattoli1999hermite}, and we address the time-ordering problem by discussing the possibility of adapting for this purpose the use of expansions such as the Volterra-Neumann or Feynman-Dyson series~\cite{Gill_2017}. The paper is organized as follows: In section~\ref{section_lag}, we summarize and extend the results contained in~\cite{Dattoli_2017,dattoli1999hermite}. In Section~\ref{sec:three}, we provide a general view on the problem of time-ordering for pseudo evolution equations, while Section~\ref{sec:four} contains a number of specific examples and final comments.

\section{Laguerre derivative, Laguerre exponential and operator-ordering}\label{section_lag}

The Laguerre transform and the associated operational calculus have played a crucial role in the theory of operational calculus~\cite{ditkin1967operational}. They have offered elements of key importance within the context of the monomiality theory~\cite{Dattoli_2000} and for the study of integro-differential equations of Volterra type~\cite{Dattoli_2006}. %
We will outline the procedure allowing the merging of ordering procedures and umbral image type methods, using as a reference example the so-called \emph{``Laguerre-calculus''} along the lines of~\cite{babusci2011umbral}.

We introduce the forthcoming discussion by going back to the paradigmatic strategy sketched in the introductory section and, accordingly, fix the following specific steps:
\begin{enumerate}
\item We specialize the operators in~\eqref{eq:PEO} to
\begin{equation}
	\hat{D}_t={}_l\partial_t
		=\partial_t t\;\partial_t\,,\qquad 
		\hat O_x=\partial_x\,,
\end{equation}
where  ${}_l\partial _t$ is the \emph{Laguerre derivative}~\cite{dattoli1999hermite,Dattoli_2000,Dattoli_2005}.
\item The eigenfunction of the Laguerre derivative operator is the Bessel-like function ${}_l e(x)$~\cite{Dattoli_2018},
\begin{equation}
	{}_l e(x)=\sum_{r\geq 0}\frac{x^r}{(r!)^2}\,,
\end{equation}
which satisfies $\hat{D}_t\, {}_le(\lambda t)=\lambda\, {}_l e(\lambda t)$.
\item In view of explicit computations, it will prove advantageous to express ${}_l e(x)$ via an \emph{umbral image}~\cite{Babusci_2014,Behr_2019} (where we refer to Appendix~\ref{sec:UIT} for the explicit definition of the full formalism)
\begin{equation}\label{eq:lE1}
{}_le(x)=\fio{v e^{vx}}\,.
\end{equation}
Here, $v$ is a formal variable, and $\hat{\mathbb{I}}$ a formal integration operator, which acts according to
\begin{equation}\label{eq:fioEx}
	\fio{v^{\alpha}}:=\frac{1}{\Gamma(\alpha)}\qquad (\alpha\in \mathbb{C})\,.
\end{equation}
\end{enumerate}
We can therefore write the solution of our problem as
\begin{equation}
	F(x,t)=\fio{ve^{\alpha v t \partial_x}f(x)}
	=\fio{v f(x+\alpha v t)}\,,
\end{equation}
where we have just adopted the properties of the exponential shift operator (i.e.\ Taylor's formula). This illustrates one particularly simple scenario in which the approach sketched in the introduction may be explicitly carried out.\\

The next example addresses the problem of operator-ordering. We consider a \emph{Laguerre-type
evolution problem} (with $\hat{x}$ the operator of multiplication by $x$, i.e.\ $\hat{x}(x^n):=x^{n+1}$),
\begin{equation}
 \hat{O}_x=-(\alpha \;\hat{x}-\beta \;\partial _x)\,,
 \end{equation}
in which the novelty and the difficulty stems from the fact that it consists of the sum of two terms that are not commuting with each other (due to $[\partial_x,\hat{x}]=1\neq 0$). %
The solution of our problem can be cast in a first step into the form
\begin{equation}
	F(x,t)=\fio{v e^{-vt(\alpha\hat{x}-\beta\partial_x)}f(x)}\,.
\end{equation}
However, the PEO in this expression cannot be disentangled into the product of two exponentials, because the operators in the argument of the exponential do not commute. We thus proceed as follows:
\begin{enumerate}
\item We define the auxiliary operators
\begin{equation}
\hat{X}:=-\alpha v t \hat{x}\,,\quad \hat{Y}:=\beta v t \partial_x\,.\end{equation}
\item Applying the Weyl disentanglement rule (taking advantage of the fact that $[[\hat{X},\hat{Y}],\hat{X}]=[[\hat{X},\hat{Y}],\hat{Y}]=0$), we find that
\begin{equation}
	e^{\hat{X}+\hat{Y}}=e^{-\frac{1}{2}[\hat{X},\hat{Y}]}e^{\hat{X}}e^{\hat{Y}}\,.
\end{equation}
\item We then eventually arrive at the closed-form expression
\begin{equation}
	F(x,t)=\fio{
		ve^{-\frac{(vt)^2}{2}\alpha\beta}e^{-vt\alpha \hat{x}}e^{vt\beta \partial_x}f(x)}
		=\fio{ve^{-\frac{(vt)^2}{2}\alpha\beta}e^{-vt\alpha \hat{x}}
			f(x+v\beta t)}\,.
\end{equation}
\end{enumerate}
It important to emphasize that the operational ordering as performed above thus brings into play a  term depending on the square of the formal variable  $v$ (which commutes with the differential operators  $\hat{x}$ and $\partial _x$). Assuming for simplicity $f(x)=1$, we find that
\begin{equation}\label{eq:UIex}
	F(x,t)=\fio{ve^{-\frac{(vt)^2}{2}\alpha\beta}e^{-vt\alpha x}}\,.
\end{equation}
Finally, after Taylor-expanding the exponential and evaluating the action of the formal integral operator $\tio$ according to~\eqref{eq:fioEx}, we find the closed-form expression
\begin{equation}
	F(x,t)
	=\fio{v\sum_{n\geq 0} \frac{(-v t \alpha)^n}{n!}e^{-\frac{(vt)^2}{2}\alpha\beta}}
	=\sum_{n\geq 0}\frac{(-\alpha t)^n}{n!} 
	{}_le^{(2)}\left(-\frac{\alpha\beta t^2}{2}\right)\,,
\end{equation}
where  ${}_le_n^{(m)}(x)$ is the Bessel-like function defined as
\begin{equation}
	{}_le_n^{(m)}(x):=
	\fio{v^{n+1}e^{x\, v^m}}=
	\sum_{r\geq 0}\frac{x^r}{r!\Gamma(m r+n+1)}	\,.
\end{equation}
The example we have discussed is sufficient to demonstrate that the umbral image formalism naturally yields the solution of
evolution problems involving a Laguerre derivative and non-commuting operators.\\

In order to stress the generality and the flexibility of our method, we consider the
further example concerning the Schr\"{o}dinger-type equation
\begin{equation}\label{eq:SE}
	\imUnit \, {}_l\partial_t\Psi(x,t)=\hat{O}_x\, \Psi(x,t)\,,\quad \hat{O}_x=-\left(\alpha\,x+\tfrac{\beta}{2}\partial_x^2\right)\,.
\end{equation}
We will derive the PEO associated with eq.~\eqref{eq:SE} in complete analogy to the procedure discussed in the previous example.  As a preparatory step, let us recall for the readers' convenience the \emph{Zassenhaus formula} in its ``right- and left-oriented'' forms (see e.g.\ \cite{Casas:2012aa}), whereby for a formal variable $\lambda$ and for two composable linear operators $X$ and $Y$ one has
\begin{equation}\label{eq:Zassenhaus}
\begin{aligned}
	e^{\lambda(X+Y)}&=e^{\lambda X} e^{\lambda Y}
	e^{\lambda^2 C_2(X,Y)}e^{\lambda^3 C_3(X,Y)}
	e^{\lambda^4 C_4(X,Y)}\cdots\\
	e^{\lambda(X+Y)}&=\dotsc
	e^{\lambda^4\hat{C}_4(X,Y)}
	e^{\lambda^3\hat{C}_3(X,Y)}e^{\lambda^2\hat{C}_2(X,Y)}
	e^{\lambda Y} e^{\lambda X}\\
	C_2(X,Y)&=\tfrac{1}{2}[Y,X]\,,\quad C_3(X,Y)=\tfrac{1}{3}[C_2(X,Y),X+2Y]\\
	C_n(X,Y)&=\tfrac{1}{n!}\left(
		\tfrac{d^n}{d\lambda^n}\left(
		e^{-\lambda^{n-1}C_{n-1}(X,Y)}
			\dotsc
			e^{-\lambda^{3}C_{3}(X,Y)}
			e^{-\lambda^{2}C_{2}(X,Y)}\,
			e^{-\lambda Y}e^{-\lambda X}e^{\lambda(X+Y)}
		\right)
	\right)\bigg\vert_{\lambda\to0}\qquad (n\geq 3)\\
	\hat{C}_n(X,Y)&=(-1)^{n+1}C_n(X,Y)\qquad (n\geq 2)\,.
\end{aligned}
\end{equation}
Introducing the auxiliary operators
\begin{equation}
	\hat{A}:=\lambda \partial_x^2\,,\quad
	 \hat{B}:=\kappa \hat{x}\,,\quad \lambda:=\tfrac{\imUnit}{2}\beta v t\,,\; \kappa:=\imUnit \alpha v t\,, 
\end{equation}
we find the commutation relations
\begin{equation}\label{eq:commrel}
	\left[\hat{A},\hat{B}\right]=2\kappa \lambda \partial_x\,,\quad 
	[\hat{A},[\hat{A},\hat{B}]]=0\,,\quad 
	[\hat{B},[\hat{A},\hat{B}]]= -2\kappa^2\lambda\,,
\end{equation} 
with all higher nested commutators vanishing. Thus invoking the ``left-oriented'' form of the formula (setting $X=\hat{A}$, $Y=\hat{B}$ and $\lambda=1$) and using the commutation relations~\eqref{eq:commrel}, we obtain
\begin{equation}\label{eq:ZasComp1}
	\begin{aligned}
		e^{\hat{A}+\hat{B}}
			&=e^{\frac{1}{3!}([\hat{A},[\hat{A},\hat{B}]]+2[\hat{B},[\hat{A},\hat{B}]])}e^{\frac{1}{2}[\hat{A},\hat{B}]}e^{\hat{B}}e^{\hat{A}}\\
			&\overset{(*)}{=}e^{-\frac{2}{3}\kappa^2\lambda}e^{\kappa\lambda\partial_x}e^{\kappa\hat{x}}e^{\lambda \partial_x^2}\\
			&=e^{-\frac{2}{3}\kappa^2\lambda}
			e^{\kappa(\hat{x}+\kappa\lambda)}
			e^{\kappa\lambda\partial_x}e^{\lambda\partial_x^2}\\
			&=e^{\frac{1}{3}\kappa^2\lambda}
			e^{\kappa\hat{x}}
			e^{\kappa\lambda\partial_x}e^{\lambda\partial_x^2}
	\end{aligned}
\end{equation}
Here, in the step marked $(*)$, we have taken advantage of the \emph{Crofton-Glaisher identity}~\cite{crofton79} (see also~\cite[Eq.~(I.3.17)]{Dattoli_1997} and~\cite{Dattoli_2000}), whereby for a formal power series\footnote{Note that we provide this identity in ``operational form'', i.e.\ by using the formal multiplication operator $\hat{x}$, this expression is also valid when part of larger expressions.} $f(x)$ and for an integer-valued parameter $m\geq1$,
\begin{equation}\label{eq:CGI}
e^{\lambda \partial_x^m}f(\hat{x})=f\left(\hat{x}+m\lambda \partial_x^{m-1}\right)e^{\lambda \partial_x^m}
\end{equation}
Coincidentally, this identity also permits us to compute the action of the operational expression computed in~\eqref{eq:ZasComp1} on an initial condition $\Psi (x,0)=\varphi (x)$, resulting in\footnote{Here and throughout this paper, in expressions such as $\varphi(\hat{x}+\kappa \lambda+2\lambda \partial_x)1$, the occurrence of the symbol ``$1$'' entails that the expression is to be evaluated by expanding $\varphi(\hat{x}+\kappa \lambda+2\lambda \partial_x)$ into normal-ordered form (i.e.\ into a series in the normal-ordered monomials $\hat{x}^r\partial_x^s$ for $r,s\geq 0$), followed by acting on $1$ (which due to $\partial_x^s1=0$ for $s>0$ in effect amounts to dropping all terms of the expansion involving non-zero powers of $\partial_x$).}
\begin{equation}
	e^{\frac{1}{3}\kappa^2\lambda}
			e^{\kappa\hat{x}}
			e^{\kappa\lambda\partial_x}e^{\lambda\partial_x^2}\varphi(x)=
			e^{\frac{1}{3}\kappa^2\lambda}
			e^{\kappa\hat{x}}\varphi(\hat{x}+\kappa \lambda+2\lambda \partial_x)1\,.
\end{equation}
Combining this result with the explicit formula for the eigenfunctions ${}_le(x)$ as provided in~\eqref{eq:lE1}, we thus finally arrive at the explicit solution for $\Psi(x,t)$ evolving according to the pseudo-evolution equation~\eqref{eq:SE} with initial condition $\Psi (x,0)=\varphi (x)$:
\begin{equation}\label{eq:SEgenSol}
	\Psi(x,t)=\fio{
		ve^{\frac{1 }{6}(\imUnit vt)^3 \alpha^2\beta}
		e^{ (\imUnit vt)\alpha \hat{x}}\varphi(\hat{x}
		+\tfrac{\alpha\beta}{2}(\imUnit vt)^2+(\imUnit vt)\beta\partial_x)1
	}\,.
\end{equation}
Specializing for simplicity to the case $\varphi(x)=1$, \eqref{eq:SEgenSol} evaluates to
\begin{equation}\label{eq:PEOex3}
\begin{aligned}
\Psi(x,t)&=\fio{
		ve^{(\imUnit vt)\alpha x+\frac{1 }{6}(\imUnit vt)^3 \alpha^2\beta}
	}\,.
	\end{aligned}
\end{equation}
Coincidentally, the expression obtained in the last step has an interesting formal meaning: consider the \emph{third order Hermite polynomials} $H_n^{(3)}(x,y)$, which are defined as
\begin{equation}\label{eq:HermitePolyThirdOrder}
	H_n^{(3)}(x,y)
		:=e^{y\partial_x^3}x^n=n!\sum_{r=0}^{\left\lfloor \frac n 3\right\rfloor}
			\frac{x^{n-3r}y^r}{(n-3r)!\;r!}\,,
\end{equation}
and whose exponential generating function (EGF) reads\footnote{One may in fact derive the explicit formula for this generating function directly via use of the Crofton-Glaisher identity~\eqref{eq:CGI} combined with the results of~\eqref{eq:ZasComp1}.}
\begin{equation}\label{eq:HthreeEGF}
	\mathcal{H}^{(3)}(t;x,y):=\sum_{n\geq 0}\frac{t^n}{n!}H_n^{(3)}(x,y)
		=e^{y\partial_x^3}e^{tx}= e^{tx+t^3y}\,.
\end{equation}
We thus recognize the occurrence of the above EGF as a term in~\eqref{eq:PEOex3}, which allows us to express $\Psi(x,t)$ in the alternative form
\begin{equation}
	\Psi(x,t)=\fio{v\sum_{n\geq 0}\frac{(\imUnit vt)^n}{n!} 
	H_n^{(3)}\left(\alpha x,\tfrac{\alpha^2\beta}{6}\right)}
	=\sum_{n\geq 0}\frac{(\imUnit t)^n}{(n!)^2} H_n^{(3)}\left(\alpha x,\tfrac{\alpha^2\beta}{6}\right)\,.
\end{equation}

The examples of this introductory section have shown that a judicious combination of our suggested pseudo evolution operator (PEO) method with various elements from the theory of generalized functions and umbral image type techniques results in a toolset that allows to deal with non-standard forms of partial differential equations efficiently.\\

Before closing this section, it is worth commenting on the role played by the concepts associated with the semi-group property (or, rather, lack thereof) of the Laguerre exponential. We follow the point of view of~\cite{Dattoli_2017}, where these problems have been systematically investigated. As may be verified via an explicit calculation, one finds that the Laguerre exponential does not satisfy the semi-group property $e^{x+y}=e^x e^y$ (for $x,y$ commuting variables) of the ordinary exponential function, whence
\begin{equation}\label{eq:LaguerreNonSemiGroup}
 {}_le(x+y)\neq {}_le(x)\; {}_le(y)\,,
 \end{equation}
but rather satisfies
\begin{equation}
{}_le(x)\; {}_le(y)
	=\sum_{r\geq 0} \frac{x^r}{\left(r!\right)^2}
	\sum_{s\geq 0}\frac{y^s}{\left(s!\right)^2}
	=\sum_{n\geq 0}\frac 1{\left(n!\right)^2}\left(x{\oplus}_ly\right)^n\,.
\end{equation}
Here, the symbol  ${\oplus}_l$ denotes the \emph{composition rule}
\begin{equation}
 \left(x{\oplus}_ly\right)^n
 	:=\sum_{s=0}^n \binom{n}{s}^2 x^{n-s}y^s\,,
\end{equation}
thus yielding the so-called \emph{``Laguerre Newton binomial''}. According to the previous identities we may redefine the \emph{semi-group property} for the Laguerre exponential as
\begin{equation}\label{eq:LNP}
{}_le(x) {}_le(y)={}_le(x{\oplus}_ly)\,.
\end{equation}

The recently  introduced reformulation~\cite{Behr_2019} of the umbral calculus framework in terms of umbral image type techniques permits to understand the calculations that lead to~\eqref{eq:LNP} in a very direct manner: taking advantage of the identity (see Appendix~\ref{sec:UIT} for further details)
\begin{equation}\label{eq:fioId}
	\fio{u^{\alpha}v^{\alpha}}=\frac{\Gamma(\alpha)}{\Gamma(\alpha)}=1\quad (\alpha\in \mathbb{C}\setminus \{0,-1,-2,\dotsc\})\,,
\end{equation}
we may compute a \emph{``Laguerre Newton binomial''} type result as follows:
\begin{equation}
\begin{aligned}
{}_le(x){}_le(y)&=\fio{v_1v_2e^{v_1x+v_2y}}\\
&=\fio{u v_1v_2v_3 e^{v_3(u(v_1x+v_2y))}}\\
&=\fio{uv_1v_2{}_le(u(v_1x+v_2y))}\,.
\end{aligned}
\end{equation}
We thus indeed find that
\begin{equation}
	\fio{uv_1v_2(u(v_1x+v_2y))^n}=\sum_{r=0}^n\binom{n}{r}\frac{n! x^ry^{n-r}}{r! (n-r)!}=\left(x{\oplus}_ly\right)^n\,.
\end{equation}
This form of ``umbral image reshaping'' will prove particularly useful when considering more complicated types of special functions in the sequel.

\section{Pseudo-evolutive problems and matrix calculus}\label{sec:three}

\begin{figure}[H]
\centering
\includegraphics[scale=0.8]{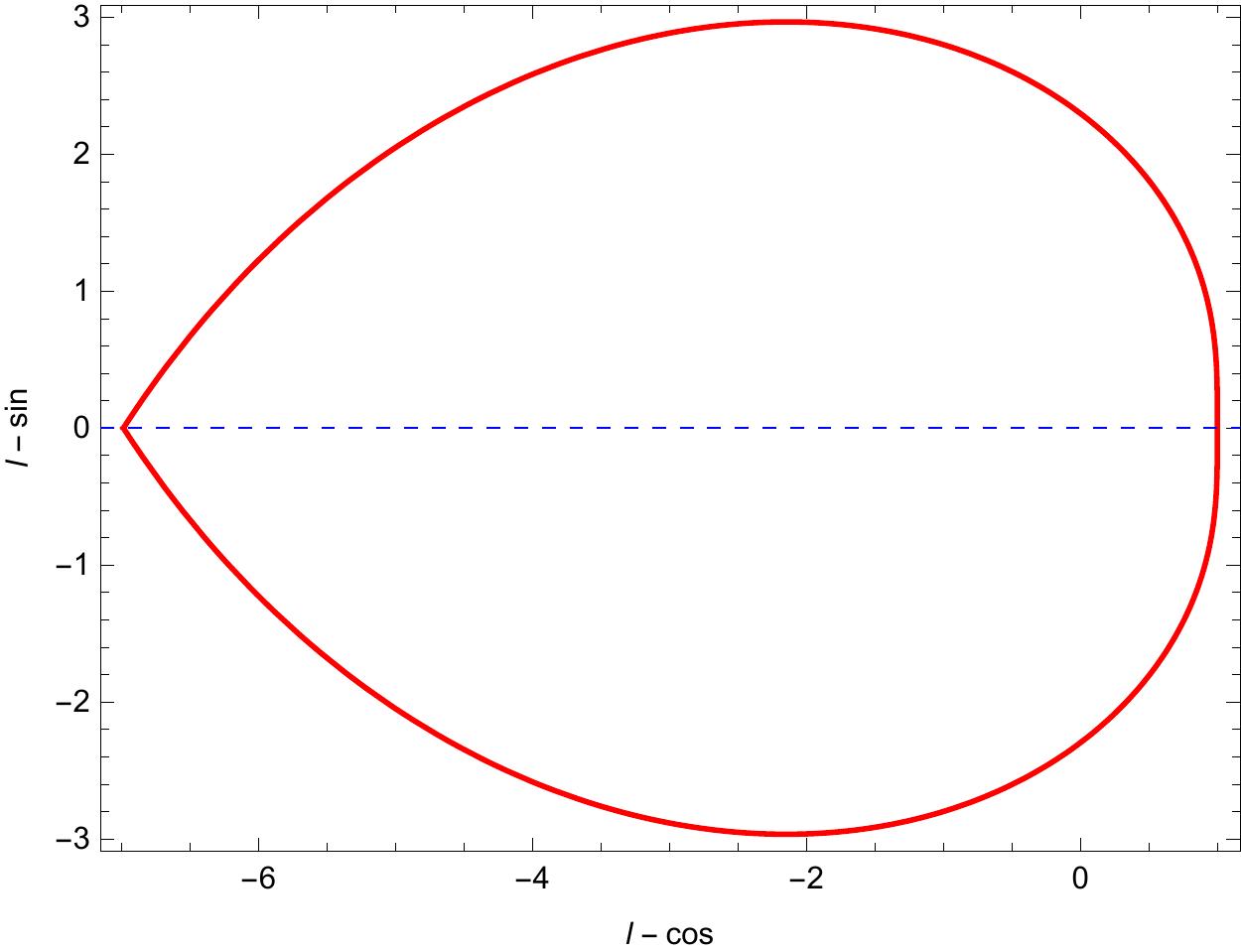}
\caption{Polar plot of the fundamental l-trigonometric relation.}\label{figsencos}
\end{figure}

\begin{figure}[H]
\centering
\includegraphics[scale=0.8]{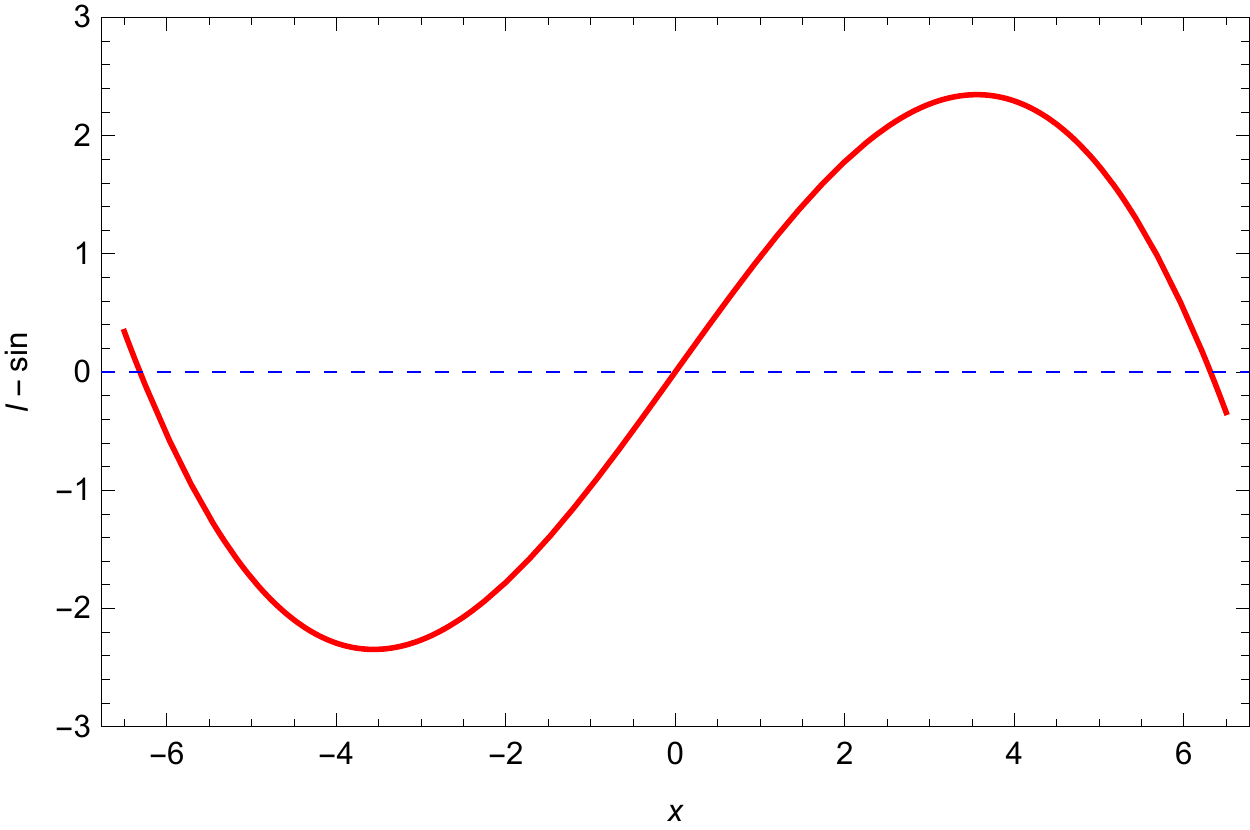}
\caption{Plot of the ${}_ls(x)$ function. }\label{figsen}
\end{figure}

In this section, we will demonstrate an extension of the previously introduced pseudo-evolution equation techniques to a form of matrix calculus. The problem we wish to address is the search of a solution for equations of the type
\begin{equation}\label{eq.32}
    {}_l\partial_t(x)\underline{Y}(t)=\hat{M}\underline{Y}(t)\,,\quad
    \underline{Y}(0)=\underline{Y}_0\,,
\end{equation}
with ${}_l\partial_t=\partial_t t \partial_t$ the Laguerre-type time-derivative, and where $\hat{M}$ and $\underline{Y}(t),\underline{Y}_0$ denote an $n\times n$ matrix and $n$-element column vectors, respectively. We specialize our discussion to the case of a non-singular $2\times 2$ matrix with eigenvalues $\lambda_\pm$. Following the paradigm of the PEO method introduced in Section~\ref{section_lag}, one may obtain a solution of~\eqref{eq.32} in the form
\begin{equation}
\begin{aligned}
    \underline{Y}(t)&=\hat{U}(t)\underline{Y}_0\,,\qquad 
    \hat{U}(t)={}_le(\hat{M}t)=\sum_{n=0}^{\infty}\frac{(\hat{M}t)^n}{(n!)^2}\,.
\end{aligned}
\end{equation}
By application of the Cayley-Hamilton theorem, we thus obtain
\begin{equation}\label{eq.34}
    {}_le(\hat{M}t)=\frac{1}{\lambda_+-\lambda_-}\Big[(\lambda_+\hat{1}-\hat{M}){}_le(\lambda_-t)-(\lambda_-\hat{1}-\hat{M}){}_le(\lambda_+ t))\Big]\,,
\end{equation}
where $\hat{1}$ denotes the $2\times2$ unit matrix. Let us now consider for illustration a matrix with zero diagonal entries and imaginary eigenvalues, namely
\begin{equation}
   \hat{M}=\begin{pmatrix} 0 & -\alpha \\ \beta & 0 \end{pmatrix}\,. 
\end{equation}
According to~\eqref{eq.34}, the corresponding PEO $\hat{U}(t)$ can be reduced to the ``pseudo rotation matrix''
\begin{equation}
      {}_le(\hat{M}t)=\begin{pmatrix}    {}_lc(\sqrt{\alpha\beta}t) & -\sqrt{\frac{\alpha}{\beta}}   {}_ls(\sqrt{\alpha\beta}t) \\ \sqrt{\frac{\beta}{\alpha}}{}_ls(\sqrt{\alpha\beta}t) &  {}_lc(\sqrt{\alpha\beta}t) \end{pmatrix}\,,
\end{equation}
where ${}_lc(x)$ and ${}_ls(x)$ denote the \emph{Laguerre (co-)sine  functions}~\cite{Dattoli_2018} defined as
\begin{equation}\label{eq:LaguerreCoSine}
     {}_lc(x)=\frac{ {}_le(\imUnit x)+{}_le(-\imUnit x)}{2}\,,\quad 
  {}_ls(x)=\frac{ {}_le(\imUnit x)-{}_le(-\imUnit x)}{2\imUnit}\qquad (\imUnit^2=-1)\,.
\end{equation}
The relevant geometrical meaning differs from that of the ordinary circular functions and is illustrated in Figure~\ref{figsencos}, where we have plotted ${}_ls(x)$ against ${}_lc(x)$ in the region of the first negative and positive zeros of the Laguerre sine function (cf.\  Figure~\ref{figsen}).

The functions in~\eqref{eq:LaguerreCoSine} can be recognized as Bessel type functions (in particular as Kelvin ber, bei functions), and they satisfy the differential equations
\begin{equation}
     {}_l\partial^2_t {}_lc(\omega t)=-\omega^2  {}_lc(\omega t)\,,\quad 
     {}_l\partial^2_t {}_ls(\omega t)=-\omega^2  {}_ls(\omega t)\,.
\end{equation}
We also note that the Laguerre derivative satisfies the identity
\begin{equation}\label{eq.39}
    {}_l\partial^{\alpha}_t=\partial^{\alpha}_t t^\alpha \partial^\alpha_t\qquad (\alpha\in\mathbb{R})\,.
\end{equation}

Our next example illustrates a possible generalization of the PEO methods in a different direction, namely in the form of \emph{fractional} evolutive problems such as
\begin{equation}\label{eq.40}
    \partial_t^\mu\underline{Y}(t)=\hat{M}\underline{Y}(t)+\frac{t^{-\mu}}{\Gamma(1-\mu)}\underline{Y}_0\,,\quad
    \underline{Y}(0)=\underline{Y}_0\qquad (0<\mu<1)\,.
\end{equation}
Following the paradigm of the PEO method, the first step in solving~\eqref{eq.40} consists in finding an eigenfunction of the fractional differential operator $\hat{D}_t=\partial_t^{\mu}$. To this end, we recall the definition of the so-called \emph{Mittag-Leffler function (ML-f)}~\cite{mittag1903nouvelle}  $E_{\alpha,\beta}(x)$ (for $\alpha,\beta\in \bC$),
\begin{equation}\label{eq:MLf1}
	E_{\alpha,\beta}(x)
	=\sum_{r\geq 0}
	\frac{x^r}{\Gamma(\alpha r+\beta)}
	=\fio{\frac{v^{\beta}}{1-xv^{\alpha}}}\,.
\end{equation}
Here, we have yet again taken advantage of the formal integral operator $\fio{\dotsc}$ as introduced in~\cite{Behr_2019} (compare~\eqref{eq:fioEx}). An alternative useful expression for the ML-f $E_{\alpha,\beta}(x)$ may be obtained via taking a suitable Laplace transform of~\eqref{eq:MLf1}, whence
\begin{equation}\label{eq:MLf2}
	E_{\alpha,\beta}(x)=\fio{v^{\beta} \int_0^{\infty} e^{-s}\, e^{v^{\alpha}\,x}\,ds}\,.
\end{equation}
For example, the alternative form~\eqref{eq:MLf2} for the ML-f allows to derive (via a suitable umbral image reshaping) a multiplicative law in a straightforward fashion:
\begin{equation}
\begin{aligned}
	E_{\alpha,\beta}(x)E_{\alpha,\beta}(y)
	&=\fio{
		(v_1v_2)^{\beta}\int_0^{\infty} e^{-s} \,
		e^{v_1^{\alpha} x+v_2^{\alpha} y}\,ds
	}\\
	&=\fio{v_3^{\beta} (uv_1v_2)^{\beta}
		\int_0^{\infty} e^{-s} \,
		e^{v_3^{\alpha} (u^{\alpha}(v_1^{\alpha}x+v_2^{\alpha} y))}\,ds
	}\\
	&\overset{\eqref{eq:MLf1}}{=}\fio{
	(uv_1v_2)^{\beta} 
	E_{\alpha,\beta}(
		u^{\alpha}(v_1^{\alpha}x+v_2^{\alpha} y))}
	\equiv E_{\alpha,\beta}(x\oplus_{E_{\alpha,\beta}} y)\,.
\end{aligned}
\end{equation}
Here, in the second step we have introduced two additional formal integration variables $u$ and $v_3$ and took advantage of the identity~\eqref{eq:fioId} in order to suitably ``reshape'' the umbral image type expression without changing its evaluation result. This then permits to utilize the formal variable $v_3$ in order to realize the defining equation for the ML-f according to~\eqref{eq:MLf1}, resulting in the ML-f at modified argument depending on the remaining formal integration variables as presented in the third step above. We thus conclude that the ``Mittag-Leffler binomial'' law should read
\begin{equation}
\begin{aligned}
(x\oplus_{E_{\alpha,\beta}} y)^n
&=\fio{(uv_1v_2)^{\beta} (u^{\alpha}(v_1^{\alpha}x+v_2^{\alpha} y)^n}\\
&=\sum_{r=0}^n\binom{n}{r}\fio{
	u^{\alpha n+\beta}v_1^{\alpha r+\beta}x^rv_2^{\alpha(n-r)+\beta} y^{(n-r)}
}\\
&=\sum_{r=0}^n\binom{n}{r}\frac{\Gamma(n\alpha+\beta)\, x^ry^{n-r}}{\Gamma(\alpha r+\beta)\Gamma(\alpha(n-r)+\beta)}\,.
\end{aligned}
\end{equation}

Back to the fractional pseudo evolution problem described in~\eqref{eq.40}, note that the Mittag-Leffler function $E_{\alpha,\beta}(x)$ may be utilized to construct a ``pseudo eigenfunction'' of the fractional time-derivative operator $\hat{D}_t=\partial_t^{\mu}$ as follows\footnote{Here, the last term in~\eqref{eq:PseudoEF} arises due to the action of the fractional derivative in the sense of Riemann-Liouville onto the constant term $1$ of $E_{\alpha,\beta}(\hat{M}t^{\mu})$, i.e.\ it is the contribution $\partial_t^{\mu} 1=t^{-\mu}/\Gamma(1-\mu)$ (compare~\cite{Dattoli_2017}).}:
\begin{equation}\label{eq:PseudoEF}
	\partial_t^{\mu}\,{\hat{D}_t} E_{\mu,1}(\hat{M}t^{\mu})
	=\hat{M}\, E_{\mu,1}(\hat{M}t^{\mu}) +\frac{t^{-\mu}}{\Gamma(1-\mu)}\,.
\end{equation}
This permits us to determine the solution of the fractional pseudo-evolution equation~\eqref{eq.40} in closed form as
\begin{equation}
	\underline{Y}(t)=E_{\mu,1}(\hat{M}t^\mu)\underline{Y}_0\,.
\end{equation}

As a final example of an interesting fractional pseudo evolution problem, which in a sense combines the technique of the previous example with the one presented in the beginning of this section, consider
\begin{equation}\label{eq.44}
    \partial_t^\mu\, F(x,t)
    =\alpha\hat{O}_x F(x,t)+\frac{t^{-\mu}}{\Gamma(1-\mu)}f(x)\,,\quad
    F(x,0)=f(x) \qquad (0<\mu<1)\,.
\end{equation}
Following the previously presented strategy, we find an explicit solution of~\eqref{eq.44} in the form
\begin{equation}
    F(x,t)=E_{\mu,1}(\alpha t^\mu\hat{O}_x)f(x)\,.
\end{equation}
Let us then specialize this result to the case of $\alpha=1$, and with a differential operator $\hat{O}_x$ as in~\eqref{eq:SE},
\[
	\hat{O}_x=-\left(\alpha\,x+\tfrac{\beta}{2}\partial_x^2\right)\,.
\]
Taking advantage of the form of the Mittag-Leffler function $E_{\alpha,\beta}(x)$ as presented in~\eqref{eq:MLf2},
\begin{equation}
	E_{\mu,1}(\hat{O}_x\,t^{\mu})=\fio{
		v\int_0^{\infty}e^{-s}\, e^{(vt)^{\mu}\,\hat{O}_x}\,ds
	}\,,
\end{equation}
and by performing an analysis based on the Zassenhaus formula~\eqref{eq:Zassenhaus} and the Crofton-Glaisher identity~\eqref{eq:CGI}, we obtain the general formula (for $F(x,0)=f(x)$):
\begin{equation}\label{eq:PEOfract}
\begin{aligned}
F(x,t)&=E_{\mu,1}(\alpha t^\mu\hat{O}_x)f(x)\\
&=\fio{
		v\int_0^{\infty}e^{-s}\, 
		e^{-\frac{1}{6}\alpha^2\beta (vt)^{3\mu}}
	e^{-\alpha(vt)^{\mu}\hat{x}}\, 
	f\bigg(\hat{x}+\tfrac{\alpha\beta}{2}(vt)^{\mu} 
	-\beta(vt)^{\mu}\partial_x\bigg)\,1}
\end{aligned}
\end{equation}
In particular, specializing further to the case $F(x,0)=f(x)=1$, the result of~\eqref{eq:PEOfract} may be evaluated by using the third-order Hermite polynomial exponential generating function formula~\eqref{eq:HthreeEGF}, namely
\begin{equation}\label{eq:PEOfractB}
\begin{aligned}
F(x,t)&=E_{\mu,1}(\alpha t^\mu\hat{O}_x)\,1=
\fio{
		v\int_0^{\infty}e^{-s}\, 
		e^{-\alpha (vt)^{\mu} x-\frac{1}{6}\alpha^2\beta (vt)^{3\mu}}}\\
		&\overset{\eqref{eq:HthreeEGF}}{=}
		e^{-\frac{\alpha^2\beta}{6}\partial_z^3}
		\fio{
		v\int_0^{\infty}e^{-s}\, 
		e^{(vt)^{\mu} z}}\qquad (z=-\alpha x)\\
		&\overset{\eqref{eq:MLf2}}{=}e^{-\frac{\alpha^2\beta}{6}\partial_z^3}\, E_{\mu,1}(zt^{\mu})\,.
\end{aligned}
\end{equation}
Recalling both the definition of the third-order Hermite polynomials $H^{(3)}_n(x,y)$ as given in~\eqref{eq:HermitePolyThirdOrder} and of the Mittag-Leffler function as given in~\eqref{eq:MLf1}, we may reformulate the above result in the more explicit form
\begin{equation}
	F(x,t)=\sum_{r\geq 0}\frac{t^{\mu r}}{\Gamma(\mu r+1)}H^{(3)}_r\left(-\alpha x,-\tfrac{\alpha^2\beta}{6}\right)\,.
\end{equation}

\section{Time-ordering and concluding comments}\label{sec:four}

In this section we touch upon on the problem of time-dependent pseudo-evolutive equations. We will then take a cursory look
at the possibility of extending our techniques as introduced thus far to time-ordering problems, which as will become apparent is a nontrivial
challenge even for the standard Cauchy problems.\\

To illustrate the difficulties we are going to meet, we first consider the non-homogeneous first order Laguerre differential equation
\begin{equation}
{}_l\partial _tY(t)=f(t)\,,\quad Y(0)=Y_0\,,
\end{equation}
where  $f(t)$  is a time-dependent function. The formal solution of this equation reads
\begin{equation}\label{eq:48}
Y(t)={}_l\partial _t^{-1}\left[f(t)\right]+Y_0
	=\partial_t^{-1}t^{-1}\partial_t^{-1}\left[f(t)\right]+Y_0\,,
\end{equation}
with the second identity obtained using~\eqref{eq.39}. We may then rewrite~\eqref{eq:48} into the form
\begin{equation}\label{eq:49}
Y(t)=\int_0^t\frac{dt_1}{t_1}\int_0^{t_1}f(t_2)dt_2+Y_0\,.
\end{equation}
The Laguerre integration can be carried out straightforwardly. For instance, if the integrand function is expandable as a (summable) series $f(t)=\sum_{n\geq 0}\tfrac{t^n}{n!}a_n$,  we obtain
\begin{equation}\label{eq:50}
Y(t)=\sum_{n\geq0}\frac{a_n}{n!(n+1)^2}t^{n+1}+Y_0\,,
\end{equation}
which is valid provided that we can exchange summation and integral sign, and given suitable convergence properties.\\

The problem becomes more difficult if we consider the equation
\begin{equation}\label{eq:51}
{}_l\partial _tY(t)=f(t)Y(t)\,,\quad Y(0)=Y_0\,,
\end{equation}
whose solution is obtained through a judicious application of the recipes we have discussed in the previous sections. We may indeed use the corresponding equation for the ordinary derivative, namely
\begin{equation}
	y(t)=y_0 e^{\int_0^t f(t')dt'}\,,
\end{equation}
by replacing the exponential by its Laguerre counterpart, and the relevant argument by a suitable integration of the function $f(t)$. To better illustrate this technique, let us proceed by transforming~\eqref{eq:51} into an integral equation,
\begin{equation}\label{eq:52}
Y(t)={}_l\partial_t^{-1}\left[f(t)Y(t)\right]
=\int_0^t\frac{dt_1}{t_1}\int_0^{t_1}f(t_2)Y(t_2)dt_2+Y_0\,.
\end{equation}
We then eventually apply a \emph{Volterra-Neumann expansion}, defined as
\begin{equation}\label{eq:53}
\begin{aligned}
	Y(t)&=\sum_{n\geq 0} Y_n(t)\\
	Y_0(t)&=Y_0\\
	Y_{n+1}(t)
	&=\int_0^t\frac{dt_1}{t_1}\int_0^{t_1}f(t_2)Y_n(t_2)dt_2\,.
\end{aligned}
\end{equation}
It is worth noting that the inclusion of a non-homogeneous term  ${}_l\partial _tY(t)=f(t)Y(t)+g(t)$ does not introduce any further conceptual complication, but leads to additional inessential computational details, whence we omit the discussion of this more general case for brevity.\\

We may verify the correctness of the procedure by considering the example with $f(t)=-t$ in~\eqref{eq:51}, resulting in the fractional differential equation
 \begin{equation}\label{eq:54}
 \frac{1}{t}\partial_t\, t\,\partial_t\,Y(t)=-Y(t)\,,\quad Y(0)=Y_0\,.
 \end{equation}
Noting that
\begin{equation}\label{eq:55}
 \frac{1}{t}\partial_t\, t\,\partial_t
 =\partial_{\left(\frac{t}{2}\right)^2}\left(\frac{t}{2}\right)^2\;\partial_{\left(\frac{t}{2}\right)^2}\,,
 \end{equation}
we may conclude that the solution of~\eqref{eq:54} with the initial condition $Y(0)=1$ is just the Bessel function
\begin{equation}
J_0(t)={}_le\left(-\left(\tfrac{t}{2}\right)^2\right)\,.
\end{equation}
We may then verify that the same result may be obtained by summing the series in~\eqref{eq:53} directly, noting that
\begin{equation}\label{eq:56}
\int_0^t\frac{dt_1}{t_1}\int_0^{t_1}t_2dt_2
	=\left(\frac{t}{2}\right)^2\,,
\end{equation}
which then entails that performing the Volterra-Neumann expansion~\eqref{eq:53} indeed evaluates to 
\begin{equation}\label{eq:57}
Y(t)=\fio{v e^{-v\;\left(\frac{t}{2}\right)^2}}={}_le\left(-\left(\tfrac{t}{2}\right)^2\right)\,.
\end{equation}
An entirely analogous computation permits to derive the solution of~\eqref{eq:51} with  $f(t)=-t^m$  and with initial condition $Y(0)=Y_0=1$, which reads
\begin{equation}\label{eq:58}
Y(t)={}_le\left(-\frac{t^{m+1}}{(m+1)^2}\right)\,.
\end{equation}

Unfortunately, the above procedures become considerably more complicated if  $f(t)$ is not just a monic function such as $f(t)=-t^m$ as above. This is in fact a direct consequence of the lack of the semi-group property of the Laguerre exponential (see~\eqref{eq:LaguerreNonSemiGroup}). If for instance  $f(t)=\cos(t)$, the solution of~\eqref{eq:51} with initial condition $Y(0)=Y_0=1$ indeed becomes rather intricate:
\begin{equation}
\begin{aligned}
Y(t)&=\sum_{n\geq 0} Y_n(t)\\
    Y_n(t)&=\sum_{r=0}^{\infty}
    	\frac{(-1)^r t^{2r+n}}{(2r+n)^2}{}_na_r\quad (n\geq 1)\,,\qquad Y_0(t)=Y_0=1\\
    {}_na_r&=\sum_{k=0}^{r}\frac{{}_{n-1}a_k}{(2k+n-1)^2[2(r-k)]!}\quad (n\geq2)\,,\qquad {}_1a_r=\frac{1}{(2r)!}\,.
    \end{aligned}
\end{equation}

Our formalism remains applicable when passing to the setting of fractional derivatives. Consider for illustration the fractional evolution equation
\begin{equation}\label{eq:60}
	\partial_t^{\alpha} Y(t)=f(t)Y(t)+Y_0\frac{t^{-{\alpha}}}{\Gamma(1-\alpha)}\,,\quad Y(0)=Y_0\,,
\end{equation}
which may be transformed into integral form via noting that according to the definition of the fractional derivative in the sense of Riemann-Liouville, one finds that $\partial_t^{\alpha} 1=t^{-\alpha}/\Gamma(1-\alpha)$, and thus
\begin{equation}\label{eq:61}
 Y(t)=\partial_t^{-\alpha }\left[f(t)Y(t)\right]+Y_0\,.
 \end{equation}
 The use of the Riemann-Liouville integral in order to evaluate the action of $\partial_t^{-\alpha }$ yields~\cite{Diethelm_2010} 
\begin{equation}\label{eq:62}
Y(t)=\frac{1}{\Gamma(\alpha)}\int_0^t f(\tau)Y(\tau)\,(t-\tau)^{\alpha-1}d\tau + Y_0\,,
\end{equation}
and the coefficients $Y_n(t)$ of the Volterra-Neumann expansion~\eqref{eq:53} consequently satisfy the following recursion equation (with $Y_0(\tau)=Y_0$ as before):
\begin{equation}\label{eq:63}
 Y_{n+1}(t)
 =\frac{1}{\Gamma (\alpha )}
 	\int_0^t f(\tau )Y_n(\tau )\,
	(t-\tau )^{\alpha-1}d\tau\qquad (n\geq 0)\,.
\end{equation}
Specializing for illustration to the case of  $f(t)=-t$ and $Y_0=1$,  the expansion terms evaluate to  
\begin{equation}
Y_n(t)=\left(-\frac{t^{\alpha+1}}{\Gamma(\alpha)}\right)^n\left[
	\prod_{k=0}^{n-1}B\Big(k(\alpha+1)+2,\alpha\Big)
\right]\,,
\end{equation}
where  $B(x,y)$ denotes the Euler Beta function.\\

The notions we have developed so far are in fact a necessary prerequisite for the development of the concepts associated with time-ordering. Let us thus pass to an illustrative first problem requiring explicit time-ordering, in the form of the matrix equation
\begin{equation}\label{eq:82}
\partial_t^{\alpha}\underline{Y}(t)
	=\hat{M}(t)\underline{Y}(t)+\frac{t^{-\alpha}}{\Gamma(1-\alpha)}\, \underline{Y}_0\,,\qquad
	\underline{Y}(0)=\underline{Y}_0\,.
\end{equation}
Here, $\underline{Y}(t)$ and $\underline{Y}_0$ denote $n$-column vectors, while $\hat{M}(t)$ denotes a non-singular time-dependent  $n\;\times \;n$ matrix, which is in general assumed to be non-commuting with itself at different times (i.e.\ $[\hat{M}(t),\hat{M}(t')]\neq 0$ for $t\neq t'$). It is evident that also in this case the most appropriate treatment is a series expansion, but ordering criteria such as those inherent in the well-known \emph{Dyson expansion} are necessary. We will therefore write the formal  solution of~\eqref{eq:82} as
\begin{equation}\label{eq:83}
\underline{Y}(t)=\frac{1}{\Gamma(\alpha)}\int_0^t\hat{M}(\tau)\underline{Y}(\tau)(t-\tau)^{\alpha-1}\,d\tau+\underline{Y}_0\,.
\end{equation}
The corresponding Volterra-Neumann series reads
\begin{equation}\label{eq:84}
 \underline{Y}(t)=\hat{U}(t)\underline{Y}_0\,,
 \end{equation}
with a matrix-valued evolution operator  $\hat U(t)$ defined as
\begin{equation}\label{eq:85}
\begin{aligned}
\hat{U}(t)
	&=\hat{1}
	+\frac{1}{\Gamma(\alpha)}\int_0^t\hat{M}(t_1)(t-t_1)^{\alpha-1}\,dt_2\\
	&\quad +\frac{1}{\Gamma(\alpha)^2}\int_0^t\hat{M}(t_2)(t-t_2)^{\alpha-1}\left[
\int_0^{t_2}\hat{M}(t_1)(t-t_1)^{\alpha-1}dt_1
\right]\,dt_2\\
&\quad +\frac{1}{\Gamma(\alpha)^3}\int_0^t\hat{M}(t_3)(t-t_3)^{\alpha-1}\left[
\int_0^{t_3}\hat{M}(t_2)(t-t_2)^{\alpha-1}
\left[
\int_0^{t_2}\hat{M}(t_1)(t-t_1)^{\alpha-1}dt_1
\right]\,
dt_2
\right]\,dt_3\\
&\quad +\dotsc
\end{aligned}
\end{equation}
The series in~\eqref{eq:85} has been obtained by translating to the fractional integration setting the usual expression given e.g.\ in~\cite{louisell1973quantum}. The derivation of the corresponding Feynman-Dyson series along with the associated diagrammatic interpretation will be discussed in a forthcoming investigation.\\

In this paper, we have demonstrated that the operator and time-ordering techniques familiar from the setting of ordinary differential calculus may be exploited for analyzing fractional and for Laguerre operators as well. The methods we have highlighted are based on a suitable interpretation of operators and functions in umbral image form. The price to be paid is the demand  for a certain level of abstraction allowing the search of a common thread yielding the pathway to generalized ordered formulae of Feynman-Dyson type. 


\vspace{6pt} 



\authorcontributions{conceptualization, G.D.; methodology, N.B., G.D.; validation, N.B., G.D.; formal analysis, N.B., G.D., A.L.; writing --- original draft preparation, N.B., G.D., A.L.; writing --- review and editing, N.B., A.L.}


\funding{The work of N.B. was supported by a \emph{H2020 Marie Sk\l{}odowska-Curie Actions Individual Fellowship} (grant \# 753750 -- RaSiR).}

\acknowledgments{N.B. would like to thank the LPTMC (Paris 06) and ENEA Frascati for warm hospitality.}

\conflictsofinterest{The authors declare no conflicts of interest. The founding sponsors had no role in the design of the study; in the collection, analyses, or interpretation of data; in the writing of the manuscript, and in the decision to publish the results.%
}

\appendixtitles{no} 
\appendixsections{multiple} 
\appendix
\section{The umbral image type technique of~\cite{Behr_2019}}
\label{sec:UIT}

For the readers' convenience, we briefly recall the central definition of the umbral image type technique as introduced in~\cite{Behr_2019}:

\begin{Definition}[Definition~2 of~\cite{Behr_2019}]\label{def:IT}
Let $\cA=\{\lambda\}\uplus\cU\uplus\cV\uplus \cX$ be an alphabet of formal variables, where $\uplus$ denotes the operation of disjoint union, and where $\{\lambda\}$, $\cU$, $\cV$ and $\cX$ are four (disjoint) alphabets of auxiliary formal variables. We will typically employ notations such as $\cX=\{x,y,x_1,x_2,\dotsc\}$, where we make use of the indexed variable notations in case of many variables for convenience. Let furthermore $\cA_\bullet=\cA\setminus\{\lambda\}$. 

We define a \emph{formal integration operator} $\tio$ via specifying first its domain $\dom{\tio}$ as 
\begin{equation}\label{eq:domainTIO}
\dom{\tio}:=\{S\in \bC[G_\mathbb{C}(\cA_\bullet)][[\lambda]]\mid \mbox{for all } \cA^{\alpha}\in supp(S) : \range{\alpha\vert_{\cU}}\subset\bC\setminus\bZ_{\leq 0}\}\,,
\end{equation}
whence elements of $\dom{\tio}$ are formal power series in $\lambda$ with coefficients that are generalized polynomials over the alphabet $\cA_{\bullet}$ (where the extension to formal power series requires a suitable notion of summability, see below). {Here, the notation $\range{\alpha\vert_{\cU}}\subset\bC\setminus\bZ_{\leq 0}$ entails that functions $\alpha:\cA_{\bullet}\rightarrow \bC$ are required to not take zero or negative integer  values when evaluated on elements of $\cU$.} Then for some monomial $\cA^{\alpha}\in \dom{\tio}$, which reads more explicitly (recall that by definition of $\bC^{(\cA)}$, we have that $\alpha(u_i)\neq 0$ and $\alpha(v_j)\neq 0$ for only finitely many indices $i$ and $j$)
\[
\cA^{\alpha}=u_1^{\alpha(u_1)}u_2^{\alpha(u_2)}\dotsc v_1^{\alpha(v_1)}v_2^{\alpha(v_2)}\dotsc\,,
\]
the action of $\tio$ on $\cA^{\alpha}$ is defined as 
\begin{equation}
\begin{aligned}
\tio\left(\cA^{\alpha}\right)&:=\left(\prod_{u_i \in U(\alpha)}\int_0^{\infty}du_i\; \frac{e^{-u_i}}{u_i}\right)\left[
\left(\prod_{v_j\in V(\alpha)}\frac{1}{2\pi i}\int_{\gamma}dv_j\; e^{v_j}\right)\left[
\cA^{\tilde{\alpha}}\right]\right]\\
U(\alpha)&:=\supp{\alpha}\cap \cU\,,\; V(\alpha):=\supp{\alpha}\cap \cV\\
\tilde{\alpha}(a)&:=\begin{cases}
\alpha(a) &\text{if } a\in \cA\setminus\cV\\
-\alpha(a)\quad &\text{if } a\in \cV\,.
\end{cases}
\end{aligned}
\end{equation}

We extend $\tio$ by linearity to finite sums. For infinite sums, this requires an appropriate notion of convergence. A series $\sum_{i\in I}c_i\cA^{\alpha_i}$ will be in $\dom{\tio}$ (the domain of $\tio$) if the family $\Big(c_i\tio(\cA^{\alpha_i})\Big)_{i\in I}$ is \emph{summable} in the target (in the sense of discrete summability or compact convergence for entire functions).
\end{Definition}

Intuitively, the above definition allows to define a type of transformation of formal power series in which monomials in the auxiliary variable alphabet $\cU=\{u,u',u_1,u_2,\dotsc \}$ yield occurrences of Gamma functions in the numerators (i.e.\ $\fio{u^{\alpha}}=\Gamma(\alpha)$ for $\alpha\in \bC\setminus\{0,-1,-2,\dotsc\}$), while monomials in the formal variables of the alphabet $\cV=\{v,v',v_1,v_2,\dotsc\}$ yield occurrences of reciprocal Gamma functions (i.e.\ $\fio{v^{\beta}}=1/\Gamma(\beta)$ for $\beta\in \bC$). The only ``rule'' in translating an expressions involving Gamma and reciprocal Gamma functions into the ``umbral image form'' via using the formal integration operator $\tio$ is that if one wishes to express the \emph{product} of two expressions via the operator, we have that
\begin{equation}
	\fio{S_1}\fio{S_2}=\fio{S_1S_2}
\end{equation}
if and only if the \emph{support} of the formal power series $S_1$ and $S_2$ is disjoint (i.e.\ if $S_1$ and $S_2$ do not share any of the formal variables; cf.\ Lemma~1 of~\cite{Behr_2019}). It is this ``rule'' that replaces a lot of more heuristic concepts in the traditional umbral calculus literature. To provide a quick application example, we present the following short list of expressions for illustration (where the exponents of the formal variables of the alphabet $\cU$ are constrained to only take complex values not equal to $0$ or a negative integer):
\begin{equation}
\begin{aligned}
	\fio{u^{\alpha}}&=\Gamma(\alpha)\,, \qquad & 
	\fio{v^{\beta}}&=\frac{1}{\Gamma(\beta)}\\
	\tio\left(u^{\alpha+n}v^{\alpha}\right)
	&=\frac{\Gamma(\alpha+n)}{\Gamma(\alpha)}=(\alpha)_n\,, \qquad  & 
\tio\left(u^{\beta}v^{\beta+n}\right)&=\frac{\Gamma(\beta)}{\Gamma(\beta+n)}=\frac{1}{(\beta)_n}
\end{aligned}
\end{equation}
A more complex set of examples is provided by the following expression quoted from~\cite{Behr_2019} for the generalized hypergeometric functions, illustrating further the utility of the umbral image type formalism:
\begin{equation}
\begin{aligned}
\pFq{p}{q}{\vec{\alpha}}{\vec{\beta}}{z}\equiv\pFq{p}{q}{\seq{\alpha_i}{1\leq i\leq p}}{\seq{\beta_j}{1\leq j\leq q}}{z}&:=\sum_{n\geq 0}\frac{z^n}{n!}\frac{(\alpha_1)_n\cdots (\alpha_p)_n}{(\beta_1)_n\cdots (\beta_q)_n}\\
&=
\tio\left(
\left(\prod_{i=1}^p (u_i v_i)^{\alpha_i}\right)
\left(\prod_{j=1}^q (u_{j+p} v_{j+p})^{\beta_j}\right)\;e^{z u_1\cdots u_p v_{p+1}\cdots v_{p+q}}
\right)\,.
\end{aligned}
\end{equation}

\section{An alternative to the Zassenhaus formula}
In Section~\ref{section_lag}, we have exploited the Zassenhaus formula to derive the solution of Laguerre or fractional Schr\"odinger equations, in which the operators appearing in the arguments of the respective pseudo evolution operators may be reduced to exponentials of the type
\begin{equation}\label{a_1}
    e^{\hat{X}+\hat{Y}}=e^{\hat{X}}e^{\hat{Y}}e^{-\frac{1}{2}[\hat{X},\hat{Y}]}e^{\frac{1}{3}[\hat{Y},[\hat{X},\hat{Y}]]+\frac{1}{6}[\hat{X},[\hat{X},\hat{Y}]]}\,.
\end{equation}
An alternative to the use of eq. \eqref{a_1} is provided by the following Berry-type rule \cite{berry1966diffraction}
\begin{equation}\label{a_berry}
  e^{\hat{X}+\hat{Y}}=e^{\frac{{m}^2}{12}-\frac{{m}}{2}\hat{X}^{\frac{1}{2}}+\hat{X}}e^{\hat{Y}},
\end{equation}
which is valid if 
\begin{equation}
[\hat{X},\hat{Y}]=m\hat{X}^{\frac{1}{2}}\,.
\end{equation}
To verify the correctness of~\eqref{a_berry}, let us consider the following realization of the operators $\hat{X}$ and $\hat{Y}$:
\begin{equation}
    \hat{X}=\alpha\partial^2_x \qquad \text{and}\qquad \hat{Y}=\beta \hat{x}
\end{equation}
By applying the identity~\ref{a_berry}, noting that here indeed $[\hat{X},\hat{Y}]=m\hat{X}^{\frac{1}{2}}$ (with $m=2\sqrt{\alpha}\beta$), we obtain
\begin{equation}\label{a_3}\begin{split}
    e^{\hat{X}+\hat{Y}}&=e^{\frac{{m}^2}{12}-\frac{{m}}{2}\hat{X}^{\frac{1}{2}}+\hat{X}}e^{\hat{Y}}\\&=e^{\frac{1}{3}\alpha\beta^2-\alpha\beta\partial_x+\alpha\partial^2_x}e^{\beta x}\,.
    \end{split}
\end{equation}
This is thus consistent with the result calculated via the Zassenhaus-type formula~\eqref{a_1}. A detailed proof of the Berry-type identity~\eqref{a_berry} independent of the Zassenhaus expansion may also be found in~\cite{Dattoli_1997,babusci2010lectures}.

\reftitle{References}

\end{document}